\documentclass[12pt]{article}
\usepackage{epsfig}
\usepackage{amsfonts}
\usepackage{amscd}
\usepackage{latexsym}
\usepackage{amsmath,amssymb}
\usepackage{verbatim}
\usepackage{setspace}
\usepackage{color}
\usepackage{cite}

\usepackage[textheight=9in, textwidth=6.5in, letterpaper]{geometry}

\usepackage{hyperref}
\hypersetup{
    colorlinks,
    citecolor=black,
    filecolor=black,
    linkcolor=black,
    urlcolor=black
    linktoc=all
}

\numberwithin{equation}{section}

\def\e{{\epsilon}}
\def\cs{{\cal S}}

 \def\p{\partial}
 \def\bz{{\bar z}}
 
\def\0{{(0)}}
\def\1{{(1)}}
\def\2{{(2)}}

\def\cO{{\cal O}}

\def\ci{{\mathcal I}}

\def\<{\langle }
\def\>{\rangle }

\def\pt{{\cal P T}}
\def\Q{{\cal Q}}

\newcommand{\bea}{\begin{eqnarray}}
\newcommand{\eea}{\end{eqnarray}}
\newcommand{\be}{\begin{equation}}
\newcommand{\ee}{\end{equation}}
\newcommand{\ba}{\begin{align}}
\newcommand{\ea}{\end{align}}
\def\be{\begin{equation}}
\def\ee{\end{equation}}
\def\beq{\be\begin{array}{c}}
\def\eeq{\end{array}\ee}
\def\Phi_1{E_r }

\renewcommand{\epsilon}{\varepsilon}

   \makeatletter
  \let\over=\@@over \let\overwithdelims=\@@overwithdelims
  \let\atop=\@@atop \let\atopwithdelims=\@@atopwithdelims
  \let\above=\@@above \let\abovewithdelims=\@@abovewithdelims
\renewcommand\section{\@startsection {section}{1}{\z@}%
                                   {-3.5ex \@plus -1ex \@minus -.2ex}
                                   {2.3ex \@plus.2ex}%
                                   {\normalfont\large\bfseries}}

\renewcommand\subsection{\@startsection{subsection}{2}{\z@}%
                                     {-3.25ex\@plus -1ex \@minus -.2ex}%
                                     {1.5ex \@plus .2ex}%
                                     {\normalfont\bfseries}}

\begin{document}
\begin{titlepage}
\unitlength = 1mm

\ \\
\vskip 1cm
\begin{center}

{ \LARGE {\textsc{Magnetic Corrections to the Soft Photon Theorem}}}

\vspace{0.8cm}
Andrew Strominger

\vspace{1cm}

{\it  Center for the Fundamental Laws of Nature, Harvard University,\\
Cambridge, MA 02138, USA}\\

\begin{abstract}
The soft photon theorem, in its standard form, requires corrections when the asymptotic particle states
carry magnetic charges.  These corrections are deduced using electromagnetic duality and the resulting soft formula conjectured to be exact for all abelian gauge theories. Recent work has shown that the standard soft theorem implies an infinity of 
conserved electric charges. The associated symmetries are identified as `large' electric gauge transformations. Here the magnetic corrections to the  soft theorem are shown to imply a second infinity of 
conserved magnetic charges. The associated symmetries are identified as `large' magnetic gauge transformations. The large magnetic symmetries are naturally subsumed in a complexification of the electric ones. 

   \end{abstract}

\vspace{1.0cm}

\end{center}

\end{titlepage}

\pagestyle{empty}
\pagestyle{plain}

\def\vx{{\vec x}}
\def\p{\partial}
\def\po{$\cal P_O$}

\pagenumbering{arabic}

\tableofcontents
\section{Introduction}
The soft photon theorem \cite{low54,low,bk,ggm,Weinberg:1965nx}  relates the leading infrared behavior of  scattering amplitudes with and without single soft  photon emission \be
 \langle p_{m+1},\dots  | a_{\alpha}(q) \mathcal{S}| p_1, \dots \rangle
	 =  S_0\langle p_{m+1}, \dots  |  \mathcal{S}| p_1, \dots \rangle  +\cO(q^0)
	 \label{eq:softphoton}
\ee
where  $p_k$ is  the momentum of the $k$th particle and $a_\alpha$ annihilates the momentum $q\to 0 $ photon. The soft factor $S_0$  (equation (2.1) below) has a pole in $q$ . The formula (\ref{eq:softphoton}) is exact as long as there are no magnetic monopoles among the asymptotic particles. In this paper we argue that the general form of the relation (\ref{eq:softphoton}) remains valid in the presence of monopoles, but the formula for $S_0$ is corrected. 
Electromagnetic duality transformations are used\footnote{Purely as a computational device: no duality symmetries are assumed. It would be interesting to understand the action of duality symmetries, when they do exist, on the corrected (\ref{eq:softphoton}).} to deduce the corrected form of $S_0$ (equation (2.5) below) which is conjectured to be exact. The corrected soft formula should play an important role in understanding the infrared structure of abelian gauge theories with magnetic monopoles. 

Recently it has been understood \cite{Strominger:2013lka,He:2014cra,He:2015zea,kps} that, in the absence of magnetic monopoles,  the usual soft photon theorem is the Ward  identity of an infinite-dimensional asymptotic symmetry group comprised of certain  `large' abelian gauge transformations which do not die off at infinity. The associated infinity of conservation laws equates  arbitrary moments of the electric field measured at the past of future null infinity with the antipodal moments measured at the future of past null infinity. Here we spell out the analogous magnetic story. Abelian gauge theories also have magnetic gauge symmetries 
which shift the dual magnetic potentials. When magnetic charges are present, there is an  infinite-dimensional `large'
subgroup which acts nontrivially on the $\cs$-matrix. The infinity of associated conserved charges\footnote{In the pure electric case, a symplectic form was constructed via which the charges were shown to canonically generate the symmetries \cite{He:2014cra,kps}. This analysis has not been extended to the magnetic case.} are comprised of moments of the magnetic field. The large magnetic gauge symmetries are naturally contained in a complexification of the large electric gauge symmetries.\footnote{This comes about because, in appropriate conventions, electromagnetic duality can be locally implemented on the vector potential at null infinity by rescaling and multiplication by $i$. See section 3.}  The corrected soft photon theorem is then the Ward identity of the complexified large gauge transformations. All of these symmetries are spontaneously broken and the soft photons are the Goldstone bosons. The argument can also be run backwards: the corrected soft photon theorem implies the existence of an infinite number of conserved electric and magnetic charges, together with the associated symmetries.

This paper is organized as follows. In section 2 we derive the magnetically corrected soft photon theorem. Section 3 derives the associated asymptotic symmetries and conserved charges. Some conventions are given in the appendix. 

\section{Magnetically modified soft theorem}
The leading soft factor for an $\left( m\to n+1-m \right)$-particle  scattering process defined  in (\ref{eq:softphoton}) is \cite{low54,low,bk,ggm,Weinberg:1965nx}
	\be S_0 (q,\e_\alpha ;p_k,e_k)= \sum_{k=m+1}^{n}\frac{e_k p_k \cdot  \epsilon_\alpha}{p_k \cdot q}-\sum_{k=1}^{m}\frac{e_k p_k \cdot  \epsilon_\alpha}{p_k \cdot q} .
	 \label{sdf}
\ee
Here  $ e_k={1 \over e}\int *F$ is the electric charge of the $k$th particle.  $\e_\alpha$ is the polarization vector of the soft photon whose annihilation operator $a_\alpha$ is defined in the appendix.
We wish to find the corrections to this formula required in the presence of asymptotic particles carrying magnetic charge 
$ g_k={1\over e}\int F.$ The form of the 
monopole-induced corrections are easily deduced via an electromagnetic duality transformation. This is simply a convenient field redefinition and we are not assuming any symmetry of the theory. 
Dual variables\footnote{ Appropriate to $\theta=0$ when monopoles carry no electric charge.}, denoted by a tilde,  are defined by
\bea \tilde F&=&-{4\pi \over e^2} *F, \cr \tilde e&=&{4 \pi\over e},\cr  \tilde e_k&=&{1\over \tilde e}\int *\tilde F=g_k,\cr
\tilde g_k&=&{1 \over \tilde e}\int \tilde F =-e_k. \eea
The soft photon field strength is proportional to 
\be F=dA,~~~A=e\e_\alpha  e^{iq\cdot x}.\ee
A dual soft photon vector potential and polarization is defined by 
\be\tilde F =d\tilde A= -{4\pi \over e^2}*dA,~~~~\tilde A=\tilde e \tilde \e_\alpha  e^{iq\cdot x}.\ee
 This formula defines $\tilde \e_\alpha $ up to an irrelevant shift by $q$. $\tilde \e_\alpha $ is essentially  the Hodge dual of  $\e_\alpha $ in the spatial plane transverse to the  spatial direction of photon propagation. 
 The coupling of a magnetic monopole to a photon, at soft wavelengths much larger than the monopole size, can now be  obtained by the replacement $A\to \tilde A$,  $\e_\alpha  \to \tilde \e_\alpha $ and $e_k\to \tilde e_k=g_k$.  It follows that monopoles preserve a soft relation of the form (\ref{eq:softphoton}) while correcting  the leading soft factor to\footnote{This formula also applies to incoming photons which have negative $q^0$. The only consequence of nonzero $\theta$-term ${\theta \over 32 \pi^2} \int F*F$ is that the spectrum constrains the electric charges to be of the form $e_k=en_k-{\theta e^2 \over8\pi^2}g_k$ for integer $n_k$ \cite{Witten:1979ey}. }
	\be S_0 (q,\e_\alpha ;p_k,g_k, e_k)= \sum_{k=m+1}^{n}\frac{p_k \cdot (e_k \epsilon_\alpha +g_k\tilde \e_\alpha ) }{p_k \cdot q}-\sum_{k=1}^{m}\frac{p_k \cdot (e_k \epsilon_\alpha +g_k\tilde \e_\alpha ) }{p_k \cdot q}.
	 \label{eq:sftphoton}
\ee 
Electric and magnetic charge conservation imply that $S_0$ is separately invariant under electric and magnetic gauge transformations 
$\e_\alpha  \to \e_\alpha +q$ and $\tilde \e_\alpha  \to \tilde \e_\alpha  +q$. 

We conjecture the formula (\ref{eq:sftphoton}) is exact for all abelian gauge theories.

\section{Symmetries of the $\cs$-matrix  }
In this section we describe the nontrivially acting electric and magnetic symmetries of the 
$\cs$-matrix and derive the associated Ward identities. 
For simplicity we take all charged particles to be massive. Our analysis follows closely \cite{kps}, to which we refer the interested reader for further details. 
\subsection{Preliminaries}
The Minkowski metric in retarded coordinates  reads
\be
ds^2=-dt^2+ (dx^i)^2=-du^2 -2du dr + 2r^2\gamma_{z\bz}dz d\bz,  
\ee
where $u$ is retarded time and $\gamma_{z\bz}$ is the round metric on the unit radius $S^2$ with covariant derivative $D_z$. These are related to standard  Cartesian coordinates by
\be\label{ret}
r^2=x_ix^i, \;\;\; u=t-r, \;\;\; x^i=r\hat{x}^i(z, \bz).  
\ee 
Advanced coordinates $(v,r,z, \bz)$ near past null infinity $(\mathcal{I}^-)$ are\be\label{adv}
ds^2=-dv^2+2dvdr +2r^2\gamma_{z\bz}dz d\bz ;~~~r^2=x_ix^i, \;\;\; v=t+r, \;\;\ x^i=-r\hat{x}^i(z, \bz). 
\ee
$\mathcal{I}^+ $ ($\mathcal{I}^-$) is  the null hypersurface $r= \infty$ in retarded (advanced) coordinates.  
Due to the last minus sign in (\ref{adv}) the angular coordinates on $\mathcal{I}^+$ are antipodally related to those on $\mathcal{I}^-$ so that a light ray  passing through the interior of Minkowski space reaches  the same value of $z, \bz$ at both $\mathcal{I}^+$ and $\mathcal{I}^-$. We denote the future (past) boundary of $\mathcal{I}^+$ by $\mathcal{I}^+_+$ ($\mathcal{I}^+_-$), and the future (past) boundary of $\mathcal{I}^-$ by $\mathcal{I}^-_{+}$ ($\mathcal{I}^-_{-}$).   

Near $\mathcal{I}^+$, we assume the asymptotic expansion 
\be \label{exp}
A_u=\sum_{n=1}^{\infty}\frac{A_u^{(n)}(u,z, \bz)}{r^n}, \;\;\;\;\; A_r=\sum_{n=2}^{\infty}\frac{A_r^{(n)}(u,z, \bz)}{r^n}, \;\;\;\;\; A_z=\sum_{n=0}^{\infty} \frac{A_z^{(n)}(u,z, \bz)}{r^n} 
\ee
along with similar expansions near $\mathcal{I}^-$ and for $\tilde A_\mu$.  
Near spatial infinity the field strengths are taken to obey the usual $\pt$ and Lorentz-invariant antipodal continuity condition
\be \label{matchF}
F^{(2)}_{ru}(z,\bz)|_{\mathcal{I}^+_-}=F^{(2)}_{rv}(z,\bz)|_{\mathcal{I}^-_+},
\ee
\be \label{matchB}
F^{(0)}_{z\bz}(z,\bz)|_{\mathcal{I}^+_-}=-F^{(0)}_{z\bz}(z,\bz)|_{\mathcal{I}^-_+}.
\ee
The minus sign (\ref{matchB}) arises because our advanced and retarded coordinate systems differ by a parity transformation.

\subsection{Electric charges and symmetries}

Abelian theories have an infinite number of local electric gauge symmetries under which
\be \delta A_\mu(x)=\p_\mu\e(x), ~~~~~\delta \Psi_k(x)=i\e{e_k \over e}\Psi_k(x),\ee
where $\Psi_k$ is any field or wavefunction of charge $e_k$.  One may attempt to define associated charges as various two-surface integrals of the field strength weighted by the gauge parameters. Many such charges are either trivial or not conserved. However an infinite number of non-trivial outgoing (incoming) charges on $\ci^+$ ($\ci^-$) can be associated to `large' gauge transformations on $\ci^+$ ($\ci^-$) that approach the time-independent function $\e(z,\bz)$ on $\ci$.\footnote{Note that the value of the gauge transformation at spatial infinity depends on the direction from which it is approached.}   Explicitly these are 
\bea
	Q_\varepsilon^+ &=& \frac{1}{e^2} \int_{ \mathcal{I}_-^+} d^2z  \gamma_{z\bz}   \varepsilon   F^{(2)}_{ru},\cr
Q_\varepsilon^- &=& \frac{1}{e^2} \int_{ \mathcal{I}_+^-} d^2z  \gamma_{z\bz}   \varepsilon   F^{(2)}_{rv}, \eea  where $\e(z,\bz)$  is any function on $S^2$. 
It follows immediately from (\ref{matchF}) that these charges are conserved:
\be \label{cns} Q^+_\e=Q^-_\e.\ee
Under the associated symmetry the gauge field on $\ci^+$ transforms as
\be \label{sxi} \delta_\e A^{(0)}_z(u,z,\bz)=\p_z\e(z,\bz).\ee
It follows that the large gauge symmetry is spontaneously broken and the zero modes of $A^{(0)}_z$ -- the soft photons -- are the goldstone bosons. 

It is useful to write the charges as integrals over $\ci^\pm$. Defining the outgoing, positive-helicity soft photon operator
\be\label{zmd} F^+_z\equiv \int du F^{(0)}_{uz}, \ee  integrating by parts and using  the $\ci^+$ constraint equation 
\be \label{evolution}
\p_uF_{ru}^{(2)}+D^zF^{(0)}_{uz}+D^\bz F^{(0)}_{u\bz}=0  
\ee
one finds 
\begin{align} \label{qplus}
Q_\varepsilon^+  = \frac{1}{e^2} \int_{S^2}  d^2z     \epsilon   (\p_\bz F^+_{z}+\p_z F^+_{\bz}) +\frac{1}{e^2} \int_{\mathcal{I}_+^+}d^2z \  \gamma_{z\bz}   \epsilon   F^{(2)}_{ru} \equiv Q_S^++Q_H^+.  \end{align}  
The first piece of the charge is written in terms of the soft photon operator and will be referred to as the soft charge $Q_S^+$. The second piece is proportional to  the electric fields produced by the asymptotic outgoing hard particles and will be referred to as the hard charge $Q_H^+$. 

Similar observations apply to  $\mathcal{I}^-$. The charge is given by
\begin{align} \label{qminus}
	Q_\varepsilon^-    
		&=\frac{1}{e^2} \int_{S^2}  d^2z\    \epsilon (\p_\bz F^-_{z}+\p_z F^-_{\bz})+ \frac{1}{e^2} \int_{\mathcal{I}_-^-}d^2z \  \gamma_{z\bz}  \varepsilon   F^{(2)}_{rv} \equiv Q_S^-+Q_H^-,
\end{align}  
where $F^-_z\equiv\int dv F^{(0)}_{vz}$ creates and annihilates incoming soft photons. 

The conservation law (\ref{cns}) is equivalent to the $\mathcal{S}$-matrix Ward identity:
\begin{align}
	 \<\text{out}|  \left(Q_\varepsilon^+ \mathcal{S} - \mathcal{S} Q^-_\varepsilon\right) | \text{in} \> = 0.
\end{align}
In order to facilitate later comparison with the soft theorem, we rewrite this in the form \begin{align} \label{Ward}
	 \<\text{out}|  \left(Q_S^+ \mathcal{S} - \mathcal{S} Q^-_S\right) | \text{in} \>  =  - \<\text{out}|  \left(Q_H^+ \mathcal{S} - \mathcal{S} Q^-_H\right) | \text{in} \>.
\end{align}

In  \cite{He:2014cra,kps} it was shown that, in the absence of magnetic charges, $Q^+_\e$ ($Q^-_\e$) generates large gauge symmetries on $\ci^+$ ($\ci^-$) via commutators.  $Q_S^+$, which is linear in the soft photon operator, generates the inhomogenous transformation $[Q_s^+,A^{(0)}_z]=i\p_z\e$ in (\ref{sxi}). $Q_H^+$ generates the large gauge action on the hard particles, whose charge densities are (in the massive case) generally distributed over the asymptotic $S^2$.  However the analyses in \cite{He:2014cra,kps} assumed the absence of magnetic fields at  $\ci^\pm_\pm$ and 
do not directly apply  to the present context. Moreover we expect the value of the $\theta$ angle to affect zero mode commutators and be important for such an analysis.  We expect it remains true that the  charges $Q^\pm_\e$ generate the large electric symmetries, but we will not show it here.
\subsection{Magnetic charges and symmetries}

Abelian theories also have an infinite number of local magnetic gauge symmetries under which
\be \tilde \delta_\e \tilde A_\mu(x)=\p_\mu\e(x), ~~~~~\tilde \delta_\e \Psi_k(x)=i\e{eg_k \over 4\pi }\Psi_k(x),\ee
where $\Psi_k$ is any field or wavefunction of magnetic charge $g_k$. We consider large magnetic gauge transformations, under which 
 the dual gauge field on $\ci^+$ transforms as
\be \label{mxi} \tilde \delta_\e \tilde A^{(0)}_z(u,z,\bz)=\p_z\e(z,\bz).\ee
 Conclusions parallel to those of the previous subsection apply to the magnetic case. This is obvious by working in the terms of the dual variables, which simply amounts to putting a tilde on every variable of the previous subsection. It is also useful to describe the magnetic charges in the original variables without using duality as follows.

The infinity of outgoing and incoming magnetic charges associated  to (\ref{mxi}) are \bea
	\tilde Q_{\varepsilon}^+ &=& \frac{i}{4\pi} \int_{ \mathcal{I}_-^+} d^2z    \varepsilon   F^{(0)}_{z\bz},\cr
\tilde Q_{\e}^- &=& -\frac{i}{4\pi} \int_{ \mathcal{I}_+^-} d^2z    \varepsilon   F^{(0)}_{z\bz}, \eea  where $\e(z,\bz)$  is any function on $S^2$. 
It follows immediately from (\ref{matchB}) that these charges are conserved:
\be \tilde Q^+_\e=\tilde Q^-_\e.\ee Integrating by parts and using  the Bianchi identity (instead of the constraint equation) one finds 
\bea \label{qpls}
	\tilde Q_{\e}^+   
		=\frac{i}{4\pi} \int_{S^2} d^2z\     \epsilon   (\p_\bz F^+_{z}-\p_z F^+_{\bz}) +\frac{i}{4\pi} \int_{\mathcal{I}_+^+}d^2z   \epsilon   F^{(0)}_{z\bz} &\equiv& \tilde Q_S^++\tilde Q_H^+  \cr
\tilde Q_{\e}^-   
		=\frac{i}{4\pi} \int_{S^2} d^2z\     \epsilon   (\p_\bz F^-_{z}-\p_z F^-_{\bz}) -\frac{i}{4\pi} \int_{\mathcal{I}_-^-}d^2z   \epsilon   F^{(0)}_{z\bz} &\equiv& \tilde Q_S^-+\tilde Q_H^-  		
		.  
\eea 
 The first term creates and annihilates soft photons, while the second acts on the hard asymptotic magnetically charged particles, much as in the electric case. The magnetic Ward identity is
\begin{align} \label{Wad}
	 \<\text{out}|  \left(\tilde Q_S^+ \mathcal{S} - \mathcal{S} \tilde Q^-_S\right) | \text{in} \>  =  - \<\text{out}|  \left(\tilde Q_H^+ \mathcal{S} - \mathcal{S} \tilde Q^-_H\right) | \text{in} \>. \end{align}

\subsection{Electromagnetic charges and symmetries}

It turns out that the electric and magnetic charges and symmetries combine simply into a single complexifed charge and symmetry. To see this consider the action of duality on the fields at $\ci^+$: 
\bea
   \tilde F^{(0)}_{z\bz}&=&{4\pi i \over e^2} \gamma_{z\bz}F^{(2)}_{ru}\cr \tilde F^{(2)}_{ru}&=&{4\pi i \over e^2}\gamma^{z\bz}F^{(0)}_{z\bz}\cr
    \tilde F^{(0)}_{uz}&=&{4\pi i \over e^2}F^{(0)}_{uz}\cr \tilde F^{(0)}_{u\bz}&=&-{4\pi i \over e^2}F^{(0)}_{u\bz}\eea
    and $\ci^-$:
  \bea
   \tilde F^{(0)}_{z\bz}&=&-{4\pi i \over e^2} \gamma_{z\bz}F^{(2)}_{rv}\cr \tilde F^{(2)}_{rv}&=&-{4\pi i \over e^2}\gamma^{z\bz}F^{(0)}_{z\bz}\cr
    \tilde F^{(0)}_{vz}&=&{4\pi i \over e^2}F^{(0)}_{vz}\cr \tilde F^{(0)}_{v\bz}&=&-{4\pi i \over e^2}F^{(0)}_{v\bz}\eea  
    
It is convenient to define  $\Q^\pm_\e\equiv e Q^\pm_\e+{4\pi i \over e}\tilde Q^\pm_\e$ which may be expressed
\bea \label{dsc}  \Q^+_\e
=\frac{2}{e} \int_{S^2} d^2z\     \epsilon   \p_z F^+_{\bz} +\frac{1}{e} \int_{\mathcal{I}_+^+}d^2z   \epsilon   \left(\gamma_{z\bz} F^{(2)}_{ru}- F^{(0)}_{z\bz}\right)  &\equiv& \Q_S^++\Q_H^+\cr\Q^-_\e
=\frac{2}{e} \int_{S^2} d^2z\     \epsilon   \p_z F^-_{\bz} +\frac{1}{e} \int_{\mathcal{I}_-^-}d^2z   \epsilon   \left(\gamma_{z\bz} F^{(2)}_{rv}+F^{(0)}_{z\bz}\right)  &\equiv& \Q_S^-+\Q_H^-.\eea
These complexified charges are natural because they transform simply under duality
\be \tilde \Q^\pm_\e=-i\Q^\pm_\e. \ee
The complexified Ward identity 
\begin{align} \label{Wd}
	 \<\text{out}|  \left(\Q_S^+ \mathcal{S} - \mathcal{S} \Q^-_S\right) | \text{in} \>  =  - \<\text{out}|  \left(\Q_H^+ \mathcal{S} - \mathcal{S} \Q^-_H\right) | \text{in} \>
\end{align}
then implies both (\ref{Ward}) and (\ref{Wad}). 

The structure of (\ref{dsc}) suggests that the magnetic symmetries are contained in a complexification of the electric ones. This can be made more precise. It follows from the expansion (\ref{exp}) that $F^{(0)}_{uz}=\p_uA^{(0)}_z={e^2 \over 4\pi i} \p_u\tilde A^{(0)}_z$, so that the field strengths determine the asymptotic electric and magnetic vector potentials up to a $(z,\bz)$-dependent  integration constant. It is natural to choose this constant so that 
\be \label{drt} \tilde A^{(0)}_z={4\pi i \over e^2}A^{(0)}_z,~~~\tilde A^{(0)}_\bz=-{4\pi i \over e^2}A^{(0)}_\bz. \ee
In this framework electromagnetic duality acts locally on the vector potential at $\ci$ as a $\pi \over 2$ rotation and a rescaling. 
(\ref{drt}) requires $A^{(0)}_z$ must transform under magnetic gauge transformations as
\be \tilde \delta_{ \e}A^{(0)}_z= -{ie^2 \over 4\pi}\p_z\e, ~~~~  \tilde \delta_{ \e}A^{(0)}_\bz= {ie^2 \over 4\pi}\p_\bz\e  . \ee
$\Q^+_\e$ is associated to a real electric transformation proportional to $e\e$ and an imaginary magnetic one proportional to ${4\pi i\over  e}\e$. The associated transformation of the vector potential is
\bea \label{ffl} (e\delta_\e+{4\pi i \over e}\tilde \delta_\e)A^{(0)}_z&=& 2e\p_z\e,\cr
       (e\delta_\e+{4\pi i \over e}\tilde \delta_\e)A^{(0)}_\bz&=&0.\eea
The complexified transformation acts only on the holomorphic vector potential  $A_z^{(0)}$. 
This natural complexification of the large gauge group was encountered previously  \cite{Strominger:2013lka, He:2015zea} in recasting it as a $U(1)$ Kac-Moody symmetry acting on the conformal $S^2$ at $\ci$, and is closely related to the complexification used to set $A_\bz=0$ on the boundary when recasting 3D Chern-Simons theory as 
a WZW model \cite{Witten:1988hf,Elitzur:1989nr}.

\subsection{Soft Theorem $\to$ Ward Identity}

In this section we close the loop by showing that the magnetically modified soft photon theorem (\ref{eq:softphoton}) implies the general Ward identity (\ref{Wd}). 

The outgoing soft photon theorem  can be written
\be
\lim_{\omega \to 0} \omega \langle p_{m+1},\dots  | a_{\alpha}(q) \mathcal{S}| p_1, \dots \rangle
	 =\omega  S_0 \langle p_{m+1}, \dots  |  \mathcal{S}| p_1, \dots \rangle  
	 \label{eq:softphot}
\ee
with $S_0$ given in  (\ref{eq:sftphoton}) and conventions in the appendix. 
We parameterize the photon momentum as
\be
q^\mu=\omega[1,\hat{x}(z,\bz)]\equiv \omega \hat{q}^\mu(z,\bz),  
\ee
with $\hat{x}^2=1$ and $z$ is a complex coordinate on $S^2$ as in (\ref{ret}).
The left-hand side of  (\ref{eq:softphot}) can be written in terms of the zero mode (\ref{zmd}) \begin{align}
			F^+_{ \bz}(z,\bz) &=\int du F^{(0)}_{u\bz}\equiv				- \frac{e}{8 \pi}  \partial_\bz \hat{x}^i      \lim_{\omega\rightarrow 0}  \sum_{\alpha}\left[ 
				\omega \varepsilon^{* \alpha}_i a_\alpha(\omega  \hat{x})  
				+   \omega\varepsilon^{ \alpha}_i a^\dagger_\alpha(\omega \hat{x})  \right]  \label{eq:zeromode}   
		\end{align}  by taking a weighted sum over polarizations.  Using the identity \cite{kps}
\begin{align}
			\partial_\bz \hat{x}^i(z,\bz)   \sum_\alpha \varepsilon^{*\alpha}_i   \frac{\ p_k \cdot \varepsilon_\alpha}{p_k \cdot \hat{q}(z, \bz)}
				&=\partial_\bz \log(p_k \cdot \hat{q}),  
 \end{align} 
 the soft theorem becomes 
 \begin{align}\label{zout} \<p_{m+1},\dots| F^+_{ \bz}  \mathcal{S} |p_1,\dots\>
				=~~~~~~~~~~~~~~~~~~~~~~~~~~~~~~~~~~~~~~~~~~~~~~~~~~~~~~~~~~~~~~~~~~~~~~~~~~~~~~~~~~~ \cr ~~~~~~~~~~~-\frac{e}{8 \pi}  \left(\sum_{k=m+1}^n   (e_k+ig_k)\ \partial_\bz \log(p_k \cdot \hat{q})-\sum_{k=1}^m   (e_k+ig_k)\ \partial_\bz \log(p_k \cdot \hat{q})\right) 
					\<p_{m+1},\dots|  \mathcal{S} |p_1,\dots\> .  \end{align}
For  incoming soft photons, one has
\be
\lim_{\omega \to 0} \omega \langle p_{m+1},\dots  |\mathcal{S} a^\dagger_{\alpha}(q) | p_1, \dots \rangle
	 =-\omega  S_0^* \langle p_{m+1}, \dots  |  \mathcal{S}| p_1, \dots \rangle , 
	 \label{eq:softphotonIN}  
\ee
where $S_0$ is not real in a complex basis of polarizations.  
A nearly  identical calculation then yields		
	 \begin{align}\label{zin}
			    \<p_{m+1},\dots|  \mathcal{S}F^-_{\bz}  |p_1,\dots \>
				=~~~~~~~~~~~~~~~~~~~~~~~~~~~~~~~~~~~~~~~~~~~~~~~~~~~~~~~~~~~~~~~~~~~~~~~~~~~~~~~~~~~~~~~~~~~\cr~~~\frac{e}{8 \pi} \left(\sum_{k=m+1}^n  (e_k+ig_k)\ \partial_\bz \log(p_k \cdot \hat{q}') -\sum_{k=1}^m  (e_k+ig_k)\ \partial_\bz \log(p_k \cdot \hat{q}')  \right)
					\<p_{m+1},\dots|  \mathcal{S} |p_1,\dots\>, 
		\end{align}
where  $\hat{q}' = [1,-\hat{x}^i(z,\bz)]$.	
Taking the divergence of (\ref{zout}) and (\ref{zin}), multiplying by $\e$  and integrating over the sphere, we find\footnote{A useful identity \cite{kps} is that 
$1+2D^zD_z\log (p\cdot q) ={4\pi } F^{(2)}_{ru}(z,\bz) $ where the right hand side is the retarded Lienard-Wiechert radial electric field on $\ci^+$ for a particle with unit electric charge and momentum $p$. }
	 \begin{align}\label{try}
			  \<p_{m+1},\dots|& \int_{S^2} d^2 z  \varepsilon   \p_zF^+_{ \bz}\mathcal{S} |p_1,\dots\> \notag \\&
				 =  \frac{1 }{4 } \int_{S^2} d^2z   \varepsilon \Big( [\gamma_{z\bz} F^{(2)}_{rv}+ F^{(0)}_{z\bz}]_{\ci^-_+}-[\gamma_{z\bz} F^{(2)}_{ru}-F^{(0)}_{z\bz}]_{\ci^+_+}\Big) \<p_{m+1},\dots|  \mathcal{S} |p_1,\dots\>  
	 \end{align}
	 	  \begin{align}\label{trv}
			  \<p_{m+1},\dots|& \mathcal{S} \int_{S^2} d^2 z  \varepsilon   \p_zF^-_{ \bz}|p_1,\dots\> \notag \\&
				 =   \frac{1 }{4 } \int_{S^2} d^2z   \varepsilon \Big( [\gamma_{z\bz} F^{(2)}_{ru}-F^{(0)}_{z\bz}]_{\ci^+_-}-[\gamma_{z\bz} F^{(2)}_{rv}+F^{(0)}_{z\bz}]_{\ci^-_-}\Big) \<p_{m+1},\dots|  \mathcal{S} |p_1,\dots\>.
	 \end{align}
	 Taking the difference of (\ref{try}) and (\ref{trv}) and using the continuity conditions (\ref{matchF})-(\ref{matchB}), we reproduce (\ref{Wd})
	 \begin{align}
	 	\<p_{m+1},\dots|& \Q^+_S \mathcal{S}  - \mathcal{S} \Q^-_S |p_1,\dots\> \notag  \\&
				 =  -  \frac{1}{e} \int_{S^2} d^2z   \varepsilon \Big( [\gamma_{z\bz} F^{(2)}_{ru}-F^{(0)}_{z\bz}]_{\ci^+_+}-[\gamma_{z\bz} F^{(2)}_{rv}+F^{(0)}_{z\bz}]_{\ci^-_-}\Big) \<p_{m+1},\dots|\mathcal{S} |p_1,\dots\> \\
		&=	-\<p_{m+1},\dots|\Q_H^+\mathcal{S}-\mathcal{S} \Q_H^- |p_1,\dots\>	 .
	 \end{align}
	 In conclusion, the magnetically modified soft photon theorem is the Ward identity of complexified large electromagnetic gauge transformations.

\section*{Acknowledgements}
I am  grateful  to T. Dumitrescu, D. Kapec, M. Pate, M. Schwartz and N. Seiberg for useful discussions, and also to N. Seiberg for asking an important question and to D. Kapec and M. Pate for help with some computations.  This work was supported in part by DOE grant DE-FG02-91ER40654 and the Fundamental Laws Initiative at Harvard.

\section{Appendix: Mode Expansions  }

We normalize the gauge field so that the action is  $S=-{1 \over 4 e^2}\int F^2$ with $e$ the gauge coupling. The standard mode expansion for the vector field is then \be
A_\mu(u,r,z,\bar{z})=e\sum_{\alpha}\int \frac{d^3 q}{(2\pi)^3}\frac{1}{2\omega_q}[\epsilon^{*\alpha}_\mu(\vec{q})a_\alpha(\vec{q})e^{iq\cdot x}+\epsilon_\mu^\alpha(\vec{q})a^\dagger_\alpha(\vec{q}) e^{-iq\cdot x}],  
\ee
with
\be [a_\alpha(\vec{q}),a^\dagger_\beta(\vec{q}')]=\delta_{\alpha\beta}16\pi^3\omega\delta^3(\vec q-\vec q').\ee
The large-$r$ saddle point approximation gives: 
\be
A_z^{(0)}(u,z,\bz)=-\frac{ie}{2(2\pi)^2}\p_z \hat{x}^i  \sum_{\alpha}\int_0^\infty d\omega_q[\epsilon_i^{*\alpha}a_\alpha(\omega_q \hat{x})e^{-i\omega_q u} - \epsilon_i^\alpha a_{\alpha}(\omega_q \hat{x})^\dagger e^{i\omega_q u}].   
\ee


\begin{thebibliography}{99}
\bibitem{low54} 
  F.~E.~Low,
  ``Scattering of light of very low frequency by systems of spin 1/2,''
  Phys.\ Rev.\  {\bf 96}, 1428 (1954).
  
\bibitem{low} 
  F.~E.~Low,
  ``Bremsstrahlung of very low-energy quanta in elementary particle collisions,''
  Phys.\ Rev.\  {\bf 110}, 974 (1958).
  
\bibitem{bk} 
  T.~H.~Burnett and N.~M.~Kroll,
  ``Extension of the Low soft photon theorem,''
  Phys.\ Rev.\ Lett.\  {\bf 20}, 86 (1968).

\bibitem{ggm} 
  M.~Gell-Mann and M.~L.~Goldberger,
 ``Scattering of low-energy photons by particles of spin 1/2,''
  Phys.\ Rev.\  {\bf 96}, 1433 (1954).

  \bibitem{Weinberg:1965nx} 
  S.~Weinberg,
  ``Infrared photons and gravitons,''
  Phys.\ Rev.\  {\bf 140}, B516 (1965).
  

\bibitem{kps} 
  D.~Kapec, M.~Pate and A.~Strominger,
  ``New Symmetries of QED,''
  arXiv:1506.02906 [hep-th].


  
  
  



\bibitem{He:2014cra} 
  T.~He, P.~Mitra, A.~P.~Porfyriadis and A.~Strominger,
  ``New Symmetries of Massless QED,''
  JHEP {\bf 1410}, 112 (2014)
  [arXiv:1407.3789 [hep-th]].
  \bibitem{Strominger:2013lka} 
  A.~Strominger,
  ``Asymptotic Symmetries of Yang-Mills Theory,''
  JHEP {\bf 1407}, 151 (2014)
  [arXiv:1308.0589 [hep-th]].

  \bibitem{He:2015zea} 
  T.~He, P.~Mitra and A.~Strominger,
  ``2D Kac-Moody Symmetry of 4D Yang-Mills Theory,''
  arXiv:1503.02663 [hep-th].
\bibitem{Witten:1979ey} 
  E.~Witten,
  ``Dyons of Charge e theta/2 pi,''
  Phys.\ Lett.\ B {\bf 86}, 283 (1979).
\bibitem{Witten:1988hf} 
  E.~Witten,
  ``Quantum Field Theory and the Jones Polynomial,''
  Commun.\ Math.\ Phys.\  {\bf 121}, 351 (1989).
  \bibitem{Elitzur:1989nr} 
  S.~Elitzur, G.~W.~Moore, A.~Schwimmer and N.~Seiberg,
  ``Remarks on the Canonical Quantization of the Chern-Simons-Witten Theory,''
  Nucl.\ Phys.\ B {\bf 326}, 108 (1989).
\end{thebibliography}
\end{document}